\documentclass[twocolumn,reprint,superscriptaddress,amsmath,amssymb,aps,pra]{revtex4-2}
\usepackage{units}
\usepackage{graphicx,subfigure}
\usepackage{mathrsfs}
\usepackage{bm}% bold math
\usepackage{textcomp}
\usepackage{url}
\usepackage{dsfont} 
\usepackage{braket}
\usepackage[compact]{titlesec}
\usepackage[colorlinks=true,linkcolor=magenta,citecolor=blue]{hyperref}%
\usepackage{todonotes}
\usepackage{empheq}
\usepackage{multirow}
\usepackage{hyperref}
\usepackage{graphicx}
\hypersetup{
    citecolor=blue,
    colorlinks=true,
    linkcolor=blue,
    filecolor=blue,      
    urlcolor=blue,
}
\begin{document}
\title{Phonon-phonon interaction as a control knob for multimode splitting in a two mirror parametrically assisted optomechanical cavity}
\author{Ghaisud Din}
\email{ghaisuddin34@gmail.com}
\affiliation{Ministry of Education Key Laboratory for Nonequilibrium Synthesis and Modulation of Condensed Matter, Shaanxi Province Key Laboratory of Quantum Information and Quantum Optoelectronic Devices, School of Physics, Xi’an Jiaotong University, Xi’an 710049, China}
%\author{Muqaddar Abbas}
%\email{muqaddarabbas@outlook.com}
%\affiliation{Basic Teaching and Research Department, Shenyang Urban Construction University, Shenyang, 110167, China}
%\author{Pei Zhang}
%\email{zhangpei@mail.ustc.edu.cn}
%\affiliation{Ministry of Education Key Laboratory for Nonequilibrium Synthesis and Modulation of Condensed Matter,
%Shaanxi Province Key Laboratory of Quantum Information and Quantum Optoelectronic Devices, School of Physics, Xi’an Jiaotong University, Xi’an 710049, China}
%
%\author{Pei Zhang}
%\email{zhangpei@mail.ustc.edu.cn}
%\affiliation{Ministry of Education Key Laboratory for Nonequilibrium Synthesis and Modulation of Condensed Matter, Shaanxi Province Key Laboratory of Quantum Information and Quantum Optoelectronic Devices, School of Physics, Xi’an Jiaotong University, Xi’an 710049, China}
%
\begin{abstract}
We study a driven optomechanical cavity with two movable mirrors and an intracavity optical parametric amplifier, focusing on how direct phonon-phonon coupling changes the observed normal-mode spectrum. Although the linearized system supports three hybrid modes, in the absence of direct mechanical coupling one collective mechanical mode contributes only weakly to the optical response, leading to an apparent two-peak structure. We show that finite phonon-phonon coupling mixes this weakly participating mode with the optically visible modes, making a third resonance clearly resolvable. This change appears in the mirror displacement spectrum, the output-field spectrum, and the output quadrature spectra, and becomes more pronounced with increasing coupling strength and drive power. Our results show that direct mechanical coupling provides a useful way to tune spectral visibility and multimode hybridization in two-mirror optomechanical cavity with intracavity parametric amplification.
\end{abstract}
\maketitle
\section{INTRODUCTION}
Cavity optomechanics studies the interaction between confined electromagnetic fields and mechanical motion, where radiation pressure (or related dispersive-gradient forces) couples intracavity photons to vibrational degrees of freedom \cite{RevModPhys.86.1391,kippenberg2008cavity}. This interaction has enabled precision displacement sensing and force metrology \cite{PhysRevD.23.1693,braginsky1995quantum}, and it provides a route to control mechanical motion using dynamical backaction \cite{marquardt2009optomechanics,PhysRevLett.99.093901}. With appropriately detuned driving fields, optomechanical damping can cool a mechanical resonator, and a large body of work has pushed such systems close to, and in several platforms into, the quantum regime \cite{teufel2011sideband,chan2011laser}. In parallel, theoretical frameworks based on quantum noise and linear response have clarified how measurement backaction and imprecision compete, and how these limits can be engineered by the optical environment \cite{RevModPhys.82.1155,gardiner2004quantum}.

A central concept in optomechanics is the linearization of the intrinsically nonlinear radiation-pressure interaction around a strong coherent drive \cite{PhysRevA.77.033804,PhysRevLett.99.093902}. In the resolved-sideband regime and under suitable detuning, this yields effective beam-splitter and two-mode-squeezing Hamiltonians that capture coherent state exchange and parametric amplification between optical and mechanical fluctuations \cite{PhysRevLett.99.093902}. Such linearized interactions underpin optomechanically induced transparency (OMIT), where destructive interference produces a narrow transparency window in the optical response \cite{weis2010optomechanically,PhysRevA.81.041803}. Related experiments have demonstrated slow-light and coherent interference signatures in micro- and nano-optomechanical devices \cite{safavi2011electromagnetically,weis2010optomechanically}. These interference effects are closely tied to the hybridization of optical and mechanical susceptibilities, and they provide a sensitive spectroscopic probe of coupling rates and dissipation \cite{PhysRevA.81.041803}.

In the strong-coupling regime, coherent energy exchange between optical and mechanical modes outpaces their respective decoherence rates, and the coupled system supports hybrid normal-mode excitations rather than independent resonances \cite{verhagen2012quantum}. A widely used spectral signature of this regime is normal-mode splitting (NMS), observed as a resolvable doublet in the mechanical displacement spectrum and in the optical output spectrum \cite{PhysRevLett.101.263602,groblacher2009observation}. Theoretical analyses have connected the appearance and visibility of NMS to the linearized coupling rate, the optical linewidth, and the mechanical damping, emphasizing that both coherent coupling and stability constraints must be satisfied \cite{PhysRevLett.101.263602}. Experimentally, time-domain and frequency-domain evidence of coherent photon--phonon exchange has been reported in both optical and microwave implementations \cite{verhagen2012quantum,palomaki2013coherent}. As a result, NMS and related avoided-crossing phenomena have become standard diagnostics for accessing and characterizing coherent hybridization in optomechanical devices \cite{RevModPhys.82.1155}.

Because the visibility of NMS depends on the balance among coupling, damping, and noise, considerable effort has focused on tailoring intracavity nonlinearities or engineered reservoirs to reshape the effective susceptibility \cite{PhysRevA.93.063809,PhysRevA.88.063833}. One approach introduces a parametric process inside the cavity. In particular, Huang and Agarwal studied a movable-mirror cavity containing an optical parametric amplifier (OPA) and showed that parametric gain can enhance the double-peak structure in both the mechanical and output-field spectra, thereby aiding the observation of NMS in the resolved-sideband regime \cite{PhysRevA.81.041803,PhysRevLett.101.263602}. More broadly, phase-sensitive amplification and squeezing are known to modify intracavity fluctuations and can be used to control backaction and spectral features \cite{PhysRevA.33.4033,agarwal2012quantum}. These ideas connect to a wider program of noise-interference and reservoir engineering in optomechanics, where interference between distinct fluctuation pathways can suppress backaction or reshape the response \cite{PhysRevA.93.063809}.

A second major direction is multimode optomechanics, motivated by the possibility of building networks of coupled mechanical elements and exploiting collective bright and dark supermodes \cite{PhysRevLett.107.043603,PhysRevLett.109.223601}. Systems with more than one mechanical mode coupled to a common cavity field can display interference and mode-selective damping, enabling long-lived dark modes that are weakly hybridized with the lossy optical channel \cite{massel2012multimode,PhysRevLett.112.013602}. Architectures with multiple moving boundaries or membranes add additional mechanical coordinates, such as center-of-mass and relative motion, and enable tunable avoided crossings as parameters are scanned \cite{thompson2008strong,jayich2008dispersive}. In such multimode settings, the linearized dynamics can support several hybrid modes, but their signatures need not appear with equal visibility in measured spectra. The number of clearly resolved spectral peaks is therefore determined not only by the number of eigenmodes, but also by their linewidths, spectral weights, and coupling to the measurement channel.

This point is especially relevant in two-mirror one-cavity configurations, which have been explored as a route to mediated mechanical correlations, collective cooling, and entanglement between mechanical elements \cite{PhysRevLett.98.030405}. In such systems, the cavity field couples to each movable mirror via radiation pressure, providing a common interaction bus that can generate correlated backaction forces and effective interactions in the mechanical subspace \cite{RevModPhys.86.1391}. Depending on detuning and drive conditions, the resulting dynamics can realize coherent exchange-like interactions or predominantly dissipative couplings, each producing distinct spectral signatures \cite{weis2010optomechanically}. Furthermore, when the mechanical frequencies are not identical, hybridization can be used for frequency conversion and for engineering mechanical supermodes with tunable composition \cite{PhysRevLett.108.153604,hill2012coherent}. These considerations suggest that extending NMS studies to two movable mirrors is not only natural but also useful for understanding how hybrid modes appear in experimentally accessible optomechanical spectra \cite{PhysRevLett.107.043603}.

Alongside cavity-mediated coupling, substantial progress has been made in realizing direct phonon-phonon interactions between mechanical resonators. Parametric pumping schemes can dynamically tune the mechanical coupling rate and bridge frequency mismatch, enabling coherent phonon exchange (phonon Rabi oscillations) and higher-order multi-pump processes \cite{okamoto2013coherent}. Related work has demonstrated strong, controllable coupling and time-domain coherent control in coupled mechanical systems, establishing a mechanical analogue of nonlinear optical wave mixing \cite{faust2013coherent}. From a theoretical perspective, time-periodic modulation naturally invites a Floquet description, where effective interactions and mode structure can be engineered by modulation amplitude and phase \cite{RevModPhys.89.011004,PhysRevX.5.031011}. In optomechanical contexts, periodic driving has also been used to explore synthetic gauge fields and engineered band structures, underscoring how modulation can reshape hybrid photon-phonon spectra \cite{PhysRevX.5.031011}. In the present context, these works provide motivation for treating the phonon-phonon term as an effective and tunable coherent interaction between the two mechanical resonators.

In this work, we consider a single optical cavity interacting with two nano-mechanical oscillators and include a phase-modulated phonon-phonon interaction between them. The cavity provides radiation-pressure coupling to each oscillator, producing optical-spring shifts and cavity-mediated correlations \cite{RevModPhys.86.1391}, while the phonon-phonon term introduces a direct mechanical mixing pathway whose strength and phase can modify the observable response. Our goal is not to claim the creation of an additional eigenmode in a system that already contains three coupled linearized modes. Instead, we study how direct mechanical coupling changes the participation and spectral visibility of these hybrid modes in experimentally relevant observables. In particular, we show that when the direct mechanical coupling is absent, one collective mechanical mode can remain weakly visible in the measured spectra under the chosen parameters, leading to an apparent two-peak structure. When finite phonon-phonon coupling is introduced, this weakly participating mode becomes more clearly mixed into the observable response, and a third resonance becomes resolvable in the mechanical displacement spectrum, the output-field spectrum, and the output quadratures. In this way, direct mechanical coupling acts as a useful control parameter for tuning spectral visibility and multimode hybridization in a two-mirror optomechanical cavity assisted by an intracavity OPA.

 \begin{figure}
        \centering
        \includegraphics[width=1\linewidth]{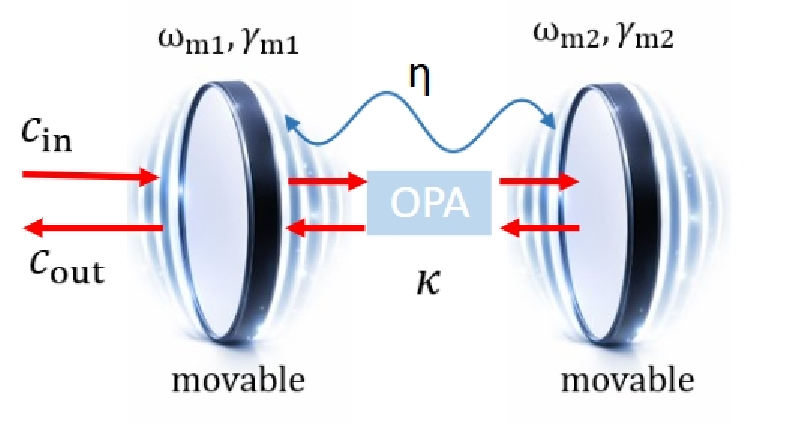}
        \caption{Schematic of the proposed two mirror OPA-assisted optomechanical system.
A driven  cavity contains an intracavity optical parametric amplifier (OPA). Both cavity mirrors are movable and behave as vibrational modes coupled to the intracavity field via radiation pressure. In addition, the two mechanical modes interact directly through a phonon-phonon coupling channel of strength $\eta$, and the optical response is controlled by phase modulation $\phi$. The cavity is pumped by a coherent laser of power $P$.}
        \label{figure1}
    \end{figure}

\section{SYSTEM AND HAMILTONIAN}
The Hamiltonian for the system shown in Fig.\ref{figure1} in a frame rotating at the laser frequency $\omega_L$ can be written as
\begin{eqnarray}
&H&=\hbar\Delta_c\eta_c-\hbar\omega_{m1}\chi \eta_c Q_{1}-\hbar\omega_{m2}\chi \eta_c Q_{2}\nonumber\\[6pt]&&+\frac{\hbar\omega_{m1}}{4}(Q_{1}^2+P_{1}^2)+\frac{\hbar\omega_{m2}}{4}(Q_{2}^2+P_{2}^2)\nonumber\\[6pt]&&+\frac{\hbar\eta}{2}[\text{sin}\phi(P_{1}Q_{2}-P_2Q_1)]+\frac{\hbar\eta}{2}[\text{cos}\phi(P_{1}P_{2}+Q_1Q_2)]\nonumber\\[6pt]&&
+i\hbar G(e^{i\theta}c^{\dagger 2}-e^{-i\theta}c^{2})+i\hbar\varepsilon(c^\dagger-c)
\label{eq:e1}
\end{eqnarray}

Here $\Delta_c=\omega_c-\omega_L$, is the detuning. $Q_j$ and $P_j$ (j=1,2) are the dimensionless position and momentum operators for the movable mirrors, defined by
$Q_j=\bigl(\sqrt{2m\omega_{mj}/\hbar}\bigr)\,q_j$
and
$P_j=\bigl(\sqrt{2/(m\hbar\omega_{mj})}\bigr)\,p_j$
with $[Q,P]=2i$.
In Eq.\ref{eq:e1}, the first term is the energy of the cavity field,
$\eta_c=c^\dagger c$ is the number of photons inside the cavity, and $c$ and $c^\dagger$ are the annihilation and creation operators for the cavity field satisfying the commutation relation $[c,c^\dagger]=1$.
The second and third term comes from the coupling of the movable mirror to the cavity field via radiation pressure; the dimensionless parameter
\begin{equation}
\chi=(\frac{1}{\omega_{mj}})(\frac{\omega_c}{L})\sqrt{\frac{\hbar}{2m\omega_{mj}}}
\label{eq:e2}
\end{equation}
is the optomechanical coupling constant between the cavity and the movable mirrors.
The fourth and fifth term corresponds to the energy of the movable mirrors. The sixth and seventh term represents the interaction between the two movable mirrors with a phonon-phonon coupling strength $\eta$ and a phase modulation $\phi$.
The secondlast term is the coupling between the OPA and the cavity field; $G$ is the nonlinear gain of the OPA, and $\theta$ is the phase of the field driving the OPA. The parameter $G$ is proportional to the pump driving the OPA. The last term describes the coupling between the input laser field and the cavity field; $\varepsilon$ is related to the input laser power $\mathcal{P}$ by
$\varepsilon=\sqrt{2\kappa\mathcal{P}/\hbar\omega_L}$, where $\kappa$ is the cavity decay rate.

Using the Heisenberg equations of motion and adding the corresponding damping and noise terms, we obtain the quantum Langevin equations as follows

\begin{eqnarray}
   \dot Q_1 &=\omega_{m1}P_1+\eta(\text{cos}\phi P_2+\text{sin}\phi Q_2) 
   \label{eq:e3}
\end{eqnarray}
\begin{eqnarray}
  \dot P_1 &=&2\omega_{m1}\chi\eta_c-\omega_{m1}Q_1-\gamma_{m1}P_1+\eta\text{sin}\phi P_2 \nonumber\\[6pt]&&-\eta\text{cos}\phi Q_2+\xi_1
  \label{eq:e4}
\end{eqnarray}
\begin{eqnarray}
   \dot Q_2 &=\omega_{m2}P_2+\eta(\text{cos}\phi P_1-\text{sin}\phi Q_1) 
   \label{eq:e5}
\end{eqnarray}
\begin{eqnarray}
  \dot P_2 &=&2\omega_{m2}\chi\eta_c-\omega_{m2}Q_2-\gamma_{m2}P_2-\eta\text{sin}\phi P_1 \nonumber\\[6pt]&&-\eta\text{cos}\phi Q_1+\xi_2
  \label{eq:e6}
\end{eqnarray}
\begin{eqnarray}
   \dot c &=&=-\text{i}(\Delta_c-\omega_{m1}\chi Q_1-\omega_{m2}\chi Q_2+\kappa)c+2Ge^{\text{i}\theta}c^\dagger\nonumber\\[6pt]&&+\varepsilon+\sqrt{2\kappa}c_{in}
   \label{eq:e7}
\end{eqnarray}
\begin{eqnarray}
   \dot c^\dagger &=&=\text{i}(\Delta_c-\omega_{m1}\chi Q_1-\omega_{m2}\chi Q_2+\kappa)c^\dagger+2Ge^{-\text{i}\theta}c\nonumber\\[6pt]&&+\varepsilon+\sqrt{2\kappa}c_{in}^\dagger
   \label{eq:e8}
\end{eqnarray}
\begin{figure*}
        \centering
        \includegraphics[width=1\linewidth]{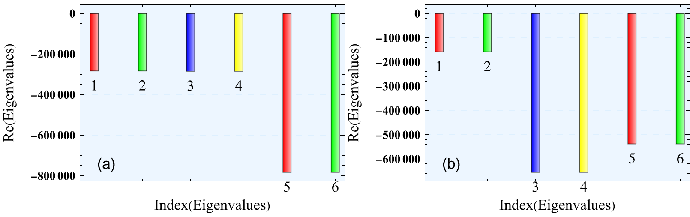}
        \caption{Stability analysis from the drift-matrix eigenvalues. Real parts of the eigenvalues of the linearized drift matrix $A$, plotted versus the eigenvalue index $i$, illustrating the stability of the steady state for two different phonon-phonon coupling strengths. (a) $\eta=0.15\,\omega_m$. (b) $\eta=0.3\,\omega_m$. In both panels, stable operation corresponds to $\text{Re}(\lambda_i)<0$ for all eigenvalues. Other parameters are $\lambda=1064\ \text{nm}$, $L=25\ \text{mm}$, $m=145\ \text{ng}$, $\kappa/2\pi=215\times10^3\text{Hz}$, $\omega_m/2\pi=947\times10^3\text{Hz}$, $T=300\ \text{mK}$, mechanical quality factor $Q^{'}=\omega_m/\gamma_m=6700$. $\theta=\pi/4$, $\phi=\pi/2$, $G=1.2\kappa$. }
        \label{figure2}
    \end{figure*}
    \begin{figure*}
        \centering
        \includegraphics[width=1\linewidth]{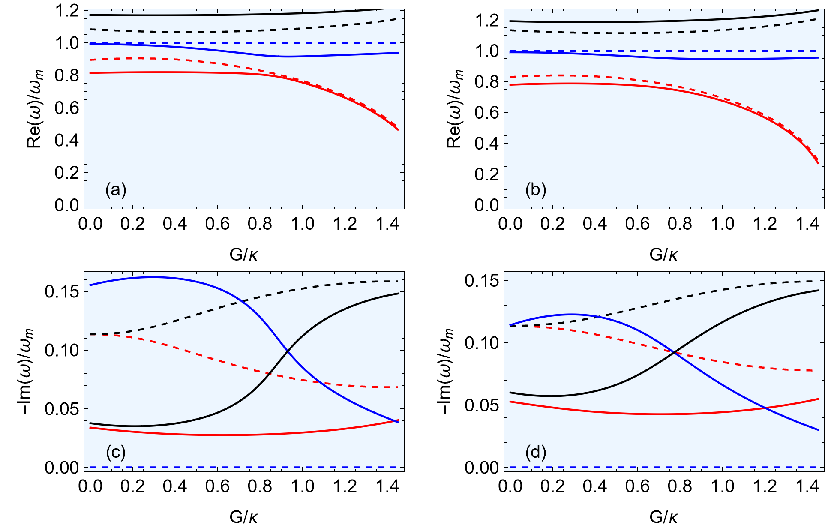}
        \caption{Effect of phonon-phonon coupling on the real and imaginary root trajectories of $\mathcal{D}(\omega)$ as a function of the parametric gain. (a,b) Real parts, Dashed curves: $\eta=0$. Solid curves: $\eta=0.15\,\omega_m$. Panel (a) is plotted for $P=3.2\,\mathrm{mW}$, while panel (b) uses the higher power $P=5.2\,\mathrm{mW}$. 
(c,d) Imaginary parts $-\text{Im}(\omega)$ of the same three branches as functions of $G/\kappa$ for the corresponding parameter sets: (c) $P=3.2\,\mathrm{mW}$ and (d) $P=5.2\,\mathrm{mW}$, comparing $\eta=0$ (dashed) with $\eta=0.15\,\omega_m$ (solid). The mechanical coupling shifts the branches and lifts the weakly participating mode into the coupled dynamics, leading to an additional, clearly resolved branch in both the real and imaginary parts of the roots. Other parameters are the same as Fig.\ref{figure2}}
        \label{figure3}
    \end{figure*}
    \begin{figure*}
        \centering
        \includegraphics[width=1\linewidth]{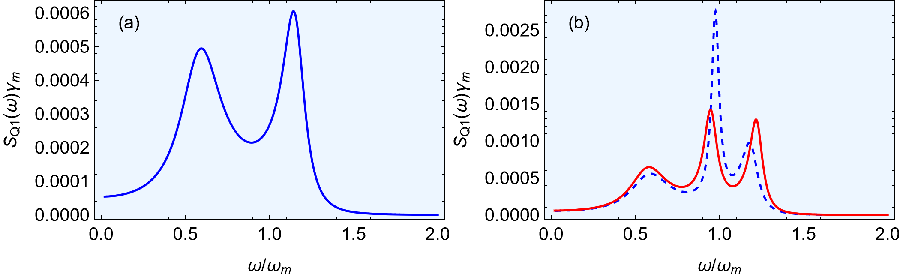}
        \caption{Position-fluctuation spectrum of one of the movable mirror and emergence of a third resonance. Normalized displacement spectrum $S_{Q_1}(\omega)$ scaled with $\gamma_m$ versus $\omega/\omega_m$ for fixed OPA gain $G=1.2\,\kappa$ and input power $P=5.2\,\mathrm{mW}$. (a) For $\eta=0$ (blue curve), the spectrum exhibits two resonant peaks. (b) Turning on phonon-phonon coupling generates an additional feature: a third peak appears for $\eta=0.1\,\omega_m$ (blue dashed curve) and becomes sharper and more clearly resolved when the coupling is increased to $\eta=0.15\,\omega_m$ (red solid curve).}
        \label{figure4}
    \end{figure*}
    \begin{figure*}
        \centering
        \includegraphics[width=0.95\linewidth]{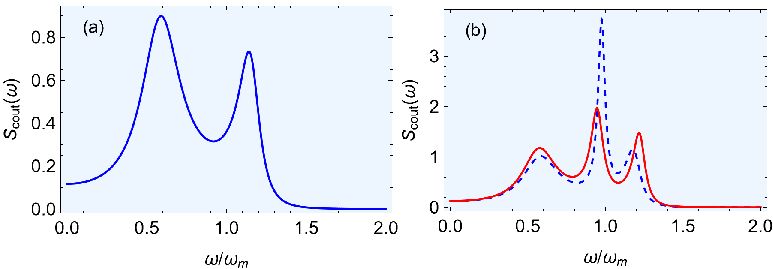}
        \caption{Output-field spectrum $S_{c{\mathrm{out}}}(\omega)$ versus $\omega/\omega_m$ for the same operating point as Fig.~4 ($G=1.2\,\kappa$ and $P=5.2\,\mathrm{mW}$). (a) In the absence of phonon-phonon coupling ($\eta=0$, blue curve), two peaks are observed in the output spectrum. (b) When mechanical coupling is introduced, a third resonance emerges for $\eta=0.1\,\omega_m$ (blue dashed curve) and becomes more pronounced for $\eta=0.15\,\omega_m$ (red solid curve).}
        \label{figure5}
    \end{figure*}
    \begin{figure*}
        \centering
        \includegraphics[width=0.95\linewidth]{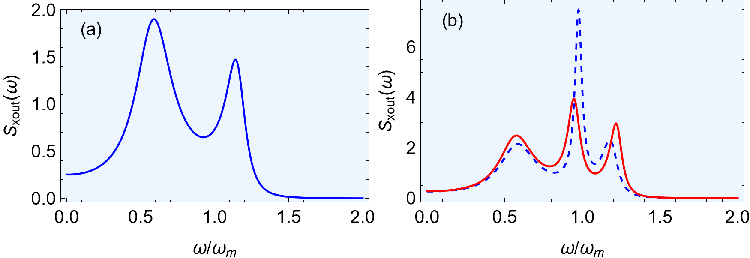}
        \caption{Spectrum of the output $x$-quadrature, $S_{x{\mathrm{out}}}(\omega)$, plotted versus $\omega/\omega_m$ for fixed OPA gain $G=1.2\,\kappa$ and input power $P=5.2\,\mathrm{mW}$. (a) For $\eta=0$ (blue curve), the spectrum shows two resonant peaks. (b) A third peak appears when phonon-phonon coupling is introduced: $\eta=0.1\,\omega_m$ (blue dashed curve) and becomes sharper and easier to resolve for $\eta=0.15\,\omega_m$ (red solid curve).}
        \label{figure6}
    \end{figure*}
    \begin{figure*}
        \centering
        \includegraphics[width=0.95\linewidth]{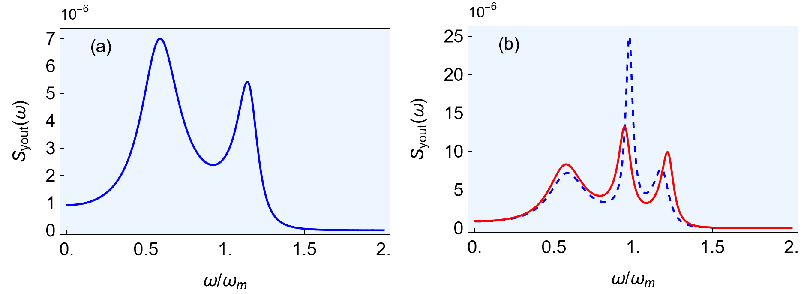}
        \caption{Spectrum of the output $y$-quadrature, $S_{y{\mathrm{out}}}(\omega)$, plotted versus $\omega/\omega_m$ for fixed OPA gain $G=1.2\,\kappa$ and input power $P=5.2\,\mathrm{mW}$. (a) For $\eta=0$ (solid curve), two peaks are observed. (b) Introducing phonon-phonon coupling produces an additional resonance: a third peak appears for $\eta=0.1\,\omega_m$ (blue dashed curve) and becomes more clearly distinguishable for $\eta=0.15\,\omega_m$ (red solid curve).}
        \label{figure7}
    \end{figure*}
We introduced the input vacuum noise operator $c_{in}$ with zero mean value, which obeys the correlation function in the time domain \cite{gardiner2004quantum},
\begin{align}
\bigl\langle \delta c_{\mathrm{in}}(t)\,\delta c_{\mathrm{in}}^\dagger(t')\bigr\rangle &= \delta(t-t'), \nonumber\\
\bigl\langle \delta c_{\mathrm{in}}(t)\,\delta c_{\mathrm{in}}(t')\bigr\rangle
=\bigl\langle \delta c_{\mathrm{in}}^\dagger(t)\,\delta c_{\mathrm{in}}(t')\bigr\rangle &=0.
\label{eq:e9}
\end{align}

The force $\xi_i$ is the Brownian noise operator resulting from the coupling of the movable mirror to the thermal bath, whose mean value is zero, and it has the following correlation function at temperature $T$ \cite{PhysRevA.63.023812}:
\begin{equation}
\bigl\langle \xi_{i}(t)\xi_{i}(t')\bigr\rangle
=\frac{1}{\pi}\frac{\gamma_{m1}}{\omega_{m1}}\int \omega e^{-i\omega(t-t')}
\left[1+\coth\!\left(\frac{\hbar\omega}{2k_BT}\right)\right]\,d\omega
\label{eq:e10}
\end{equation}
where $k_B$ is the Boltzmann constant and $T$ is the thermal bath temperature.

Following standard methods \cite{agarwal2012quantum,vogel2006quantum,scully1997quantum}, we derive the steady-state solution from Eq.~\ref{eq:e3} to Eq.~\ref{eq:e8} by setting all the time derivatives to zero. They are
\begin{equation}
    P_{1s}=P_{2s}=0
    \label{eq:e11}
\end{equation}
\begin{equation}
    Q_{1s}=\frac{2\omega_{m2}\omega_{m1}\chi \lvert c_s\rvert^2-2\eta\text{cos}\phi\omega_{m2}\chi \lvert c_s\rvert^2} {\omega_{m1}\omega_{m2}-\eta^2\text{cos}^2\phi}
    \label{eq:e12}
\end{equation}
\begin{equation}
    Q_{2s}=\frac{2\omega_{m1}\omega_{m2}\chi \lvert c_s\rvert^2-2\eta\text{cos}\phi\omega_{m1}\chi \lvert c_s\rvert^2} {\omega_{m1}\omega_{m2}-\eta^2\text{cos}^2\phi}
    \label{eq:e13}
\end{equation}
\begin{equation}
c_s=\frac{\kappa-i\Delta+2Ge^{i\theta}}{\kappa^2+\Delta^2-4G^2}\,\varepsilon
\label{eq:e14}
\end{equation}
where
\begin{equation}
\Delta=\Delta_c-\omega_{m1}\chi Q_{1s}-\omega_{m2}\chi Q_{2s}
\label{eq:e15}
\end{equation}
is the effective cavity detuning, depending on $Q_{is}$.
The $Q_{is}$ denotes the new equilibrium position of the movable mirrors relative to that without the driving field. Further $c_s$ represents the steady-state amplitude of the cavity field.

To study the normal-mode splitting of the movable mirrors and the output field, we evaluate the system’s fluctuations. Because the dynamics are nonlinear, we assume the nonlinearity is sufficiently weak, allowing us to analyze only small deviations about the steady state. Accordingly, each system operator is expressed as the sum of its steady-state expectation value and a zero-mean fluctuation term.
\begin{equation}
Q_{i}=Q_{is}+\delta Q_i,\qquad P_i=P_{is}+\delta P_i,\qquad c=c_s+\delta c.
\label{eq:e16}
\end{equation}

Inserting Eq.~\ref{eq:e16} into Eq.~\ref{eq:e3} to Eq.~\ref{eq:e8}, then assuming $|c_s|\gg 1$, the linearized quantum Langevin equations for the fluctuation operators take the form

\begin{eqnarray}
   \delta\dot Q_1 &=\omega_{m1}\delta P_1+\eta(\text{cos}\phi \delta P_2+\text{sin}\phi \delta Q_2)
   \label{eq:e17}
\end{eqnarray}
\begin{eqnarray}
   \delta\dot P_1 &=&2\omega_{m1}\chi(c_s^\ast\delta c+c_s\delta c^\dagger)-\omega_{m1}\delta Q_1-\gamma_{m1}\delta P_1\nonumber\\[6pt]&&+\eta\text{sin}\phi \delta P_2 -\eta\text{cos}\phi \delta Q_2+\xi_1
   \label{eq:e18}
\end{eqnarray}
\begin{eqnarray}
   \delta\dot Q_2 &=\omega_{m2}\delta P_2+\eta(\text{cos}\phi \delta P_1-\text{sin}\phi \delta Q_1)
   \label{eq:e19}
\end{eqnarray}
\begin{eqnarray}
  \delta \dot P_2 &=&2\omega_{m2}\chi(c_s^\ast\delta c+c_s\delta c^\dagger)-\omega_{m2}\delta Q_2-\gamma_{m2}\delta P_2\nonumber\\[6pt]&&-\eta\text{sin}\phi \delta P_1 -\eta\text{cos}\phi \delta Q_1+\xi_2
  \label{eq:e20}
\end{eqnarray}
\begin{eqnarray}
   \delta\dot c &=&-(\text{i}\Delta+\kappa)\delta c+\text{i}\omega_{m1}\chi c_s\delta Q_1+\text{i}\omega_{m2}\chi c_s\delta Q_2\nonumber\\[6pt]&&+2Ge^{\text{i}\theta}\delta c^\dagger+\varepsilon+\sqrt{2\kappa}c_{in}
   \label{eq:e21}
\end{eqnarray}
\begin{eqnarray}
   \delta\dot c^\dagger &=&-(-\text{i}\Delta+\kappa)\delta c^\dagger-\text{i}\omega_{m1}\chi c_{s}^*\delta Q_1-\text{i}\omega_{m2}\chi c_{s}^*\delta Q_2\nonumber\\[6pt]&&+2Ge^{-\text{i}\theta}\delta c+\varepsilon+\sqrt{2\kappa}c_{in}
   \label{eq:e22}
\end{eqnarray}

Introducing the cavity field quadratures
\begin{equation}
\delta x=\delta c+\delta c^\dagger,\qquad
\delta y=i(\delta c^\dagger-\delta c)
\label{eq:e23}
\end{equation}
and the input noise quadratures
\begin{equation}
\delta x_{\mathrm{in}}=\delta c_{\mathrm{in}}+\delta c_{\mathrm{in}}^\dagger,\qquad
\delta y_{\mathrm{in}}=i(\delta c_{\mathrm{in}}^\dagger-\delta c_{\mathrm{in}})
\label{eq:e24}
\end{equation}
Eq.~\ref{eq:e17} to Eq.~\ref{eq:e22} can be rewritten in the matrix form
\begin{equation}
\dot f(t)=A f(t)+n(t)
\label{eq:e25}
\end{equation}
in which $f(t)$ is the column vector of the fluctuations and $n(t)$ is the column vector of the noise sources. Their transposes are
\begin{align}
f(t)^T &= (\delta Q_1,\delta P_1,\delta Q_2,\delta P_2,\delta x,\delta y), \nonumber\\
n(t)^T &= \left(0,\xi_1,0,\xi_2,\sqrt{2\kappa}\,\delta x_{\mathrm{in}},\sqrt{2\kappa}\,\delta y_{\mathrm{in}}\right)
\label{eq:e26}
\end{align}
and the matrix $A$ is given by
\onecolumngrid
\begin{equation}
A=
\begin{pmatrix}
0 & \omega_{m1} & \eta\text{sin}\phi & \eta\text{cos}\phi & 0 & 0\\
-\omega_{m1} & -\gamma_{m1} & -\eta\text{cos}\phi & \eta\text{sin}\phi &\omega_{m1}\chi(c_s+c_s^\ast) & -i\omega_{m1}\chi(c_s-c_s^\ast)\\
-\eta\text{sin}\phi & \eta\text{cos}\phi & 0 & \omega_{m2} & 0 & 0\\
-\eta\text{cos}\phi & -\eta\text{sin}\phi & -\omega_{m2} & -\gamma_{m2} & \chi\omega_{m2}(c_s+c{s}^*) & -\text{i}\chi\omega_{m2}(c_s-c_{s}^*)\\
\text{i}\chi\omega_{m1}(c_s-c_{s}^*) & 0 & \text{i}\chi\omega_{m2}(c_s-c_{s}^*) & 0 & e^{-\text{i}\theta}G+e^{\text{i}\theta}G-\kappa & \text{i}e^{-\text{i}\theta}G-\text{i}e^{\text{i}\theta}G+\Delta\\
\chi\omega_{m1}(c_s+c_{s}^*) & 0 & \chi\omega_{m2}(c_s+c_{s}^*) & 0 & \text{i}(e^{-\text{i}\theta}G-e^{\text{i}\theta}G+\text{i}\Delta) & -e^{-\text{i}\theta}G-e^{\text{i}\theta}G-\kappa
\label{eq:e27}
\end{pmatrix}.
\end{equation}
\twocolumngrid
Taking Fourier transform of Eq.~\ref{eq:e17} to Eq.~\ref{eq:e22} by using
\begin{equation}
   f(t)=\frac{1}{2\pi}\int_{-\infty}^{+\infty} f(\omega)e^{-i\omega t}\,d\omega
   \label{eq:e28}
\end{equation}
\begin{equation}
f^\dagger(t)=\frac{1}{2\pi}\int_{-\infty}^{+\infty} f^\dagger(-\omega)e^{-i\omega t}\,d\omega
\label{eq:e29}
\end{equation}
where $f^\dagger(-\omega)=[f(-\omega)]^\dagger$, and considering $\omega_{m1}=\omega_{m2}=\omega_{m}$, $\gamma_{m1}=\gamma_{m2}=\gamma_{m}$ we obtain the spectrum of fluctuations in the position of the movable mirror, the output field spectrum, and the quadrature spectra as

\begin{eqnarray}
    &S_{Q1}(\omega)&=\frac{1}{\lvert \mathcal{D}(\omega)\rvert^2}[\mathcal{R}_{ad}(\omega)+\mathcal{T}_{h1}(\omega)\nonumber\\[6pt]&&+\mathcal{T}_{h2}(\omega)]
    \label{eq:e30}
\end{eqnarray}
\begin{eqnarray}
    &S_{\text{cout}}(\omega)&=\mathcal{V}_{1}^*(\omega)\mathcal{V}_{1}(\omega)+\mathcal{V}_{2}^*(\omega)\mathcal{V}_{2}(\omega)\nonumber\\[6pt]&&+\mathcal{F}^*(\omega)\mathcal{F}(\omega)
\end{eqnarray}
\begin{eqnarray}
   &S_{\text{xout}}(\omega)&= [\mathcal{V}_{1}(-\omega)+\mathcal{V}_1^*(\omega)][\mathcal{V}_{1}(\omega)+\mathcal{V}_{1}^*(-\omega)]\nonumber\\[6pt]&&\nonumber\\[6pt]&&+[\mathcal{V}_{2}(-\omega)+\mathcal{V}_2^*(\omega)][\mathcal{V}_{2}(\omega)+\mathcal{V}_{2}^*(-\omega)]\nonumber\\[6pt]&&\nonumber\\[6pt]&&+[\mathcal{J}(-\omega)+\mathcal{F}^*(\omega)][\mathcal{F}(\omega)+\mathcal{J}^*(-\omega)]
\end{eqnarray}
\begin{eqnarray}
   &S_{\text{yout}}(\omega)&=-[\mathcal{V}_{1}^*(\omega)-\mathcal{V}_1(-\omega)][\mathcal{V}_{1}^*(-\omega)-\mathcal{V}_{1}(\omega)]\nonumber\\[6pt]&&\nonumber\\[6pt]&&-[\mathcal{V}_{2}^*(\omega)-\mathcal{V}_2(-\omega)][\mathcal{V}_{2}^*(-\omega)-\mathcal{V}_{2}(\omega)]\nonumber\\[6pt]&&\nonumber\\[6pt]&&-[\mathcal{F}^*(\omega)-\mathcal{J}(-\omega)][\mathcal{J}^*(-\omega)-\mathcal{F}(\omega)]
\end{eqnarray}

Where the expressions for $\mathcal{D}(\omega)$, $\mathcal{R}_{ad}(\omega)$, $\mathcal{T}_{h1}(\omega)$, and $\mathcal{T}_{h2}(\omega)$ with their detailed derivation are defined in Appendix A, and the terms related to $S_{\text{cout}}(\omega)$, $S_{\text{xout}}(\omega)$, $S_{\text{yout}}(\omega)$ are defined in the Appendix B.

\section{NUMERICAL RESULTS}\label{RESULTS}
We use experimental parameters \cite{groblacher2009observation}
to investigate theoretically the normal-mode splitting in the system shown in Fig.~\ref{figure1}. The parameters are $\lambda=1064\ \text{nm}$, $L=25\ \text{mm}$, $m=145\ \text{ng}$, $\kappa/2\pi=215\times10^3\ \text{Hz}$, $\omega_m/2\pi=947\times10^3\ \text{Hz}$, $T=300\ \text{mK}$, and mechanical quality factor $Q^{'}=\omega_m/\gamma_m=6700$. In the high-temperature limit $K_{B}T \gg \hbar\omega_m$, we use $\coth(\hbar\omega/2K_{B}T)=2K_{B}T/\hbar\omega$.

We first assess the stability of the extended optomechanical setup containing an intracavity OPA, two movable mirrors, a phonon-phonon interaction of strength $\eta$, and a phase modulation $\phi$. By linearizing the equations of motion around the steady state, we obtain the drift matrix $A$ for the fluctuation variables. The stability criterion follows from Eq.~\ref{eq:e27}: stable operation requires that the real parts of all eigenvalues remain negative. For this reason, we plot $\mathrm{Re}(\lambda_i)$ as a function of the eigenvalue index $i$ in Fig.~\ref{figure2}. Panel (a) corresponds to $\eta=0.15\,\omega_m$, while panel (b) shows the stronger coupling case $\eta=0.3\,\omega_m$. The figure illustrates how increasing $\eta$ shifts the eigenvalue real parts and modifies the stability range of the system.

The behavior shown in Fig.~\ref{figure3} can be understood in terms of how the motion of the two mirrors is transferred to the cavity field. When $\eta=0$ (dashed curves), the mirrors do not exchange vibrations directly, so their motion can be reorganized into two collective mechanical patterns. One of these couples strongly to the intracavity field and therefore experiences appreciable optical backaction, while the other couples only weakly to the cavity response and remains dark or nearly dark under the chosen conditions. Because this weakly participating motion is only weakly affected by the cavity, it acquires little optically induced damping. Accordingly, in the linewidth plots $-\mathrm{Im}(\omega)$ in Fig.~\ref{figure3}(c) and Fig.~\ref{figure3}(d), one branch in the $\eta=0$ case starts close to zero, indicating a mode that remains predominantly mechanical and only weakly hybridized with the optical dynamics.

When a finite phonon-phonon coupling is introduced, $\eta=0.15\,\omega_m$ (solid curves), the situation changes qualitatively. The direct mechanical coupling mixes the two collective mechanical patterns, so the mode that was weakly visible at $\eta=0$ is no longer nearly isolated. As a result, it contributes more strongly to the cavity response and inherits additional damping through the optomechanical interaction and the OPA-modified cavity susceptibility. In Fig.~\ref{figure3}, this appears as the near-zero branch moving away from zero and evolving together with the other branches as $G$ is varied. The same mixing is visible in the root trajectories of Fig.~\ref{figure3}(a) and Fig.~\ref{figure3}(b): compared with the dashed curves, the solid curves are shifted, showing that $\eta$ modifies both the mode frequencies through $\mathrm{Re}(\omega)$ and the damping rates through $-\mathrm{Im}(\omega)$ across the gain range.

Increasing the input power from $P=3.2\,\mathrm{mW}$ in Fig.~\ref{figure3}(a,c) to $P=5.2\,\mathrm{mW}$ in Fig.~\ref{figure3}(b,d) strengthens the linearized optomechanical coupling by increasing the intracavity field. As a result, the hybridization becomes stronger and the difference between the $\eta=0$ and $\eta\neq 0$ cases becomes more pronounced. In particular, the shifts between the dashed and solid curves increase, and the redistribution of damping among the three branches becomes easier to resolve as $G$ is varied.

Figure~\ref{figure4} shows the position-fluctuation spectrum of the movable mirror, $S_{Q_1}(\omega)$, plotted as a function of the normalized frequency $\omega/\omega_m$ for fixed OPA gain and input power. Throughout Fig.~\ref{figure4}, we keep the parametric gain at $G=1.2\,\kappa$ and the driving power at $P=5.2\,\mathrm{mW}$. In the absence of phonon-phonon coupling ($\eta=0$), the spectrum in Fig.~\ref{figure4}(a) exhibits two clear resonant peaks, reflecting the dominant contribution of the optically active hybrid modes. This does not imply the absence of a third normal mode in the linearized system; rather, under these parameters the remaining collective mechanical mode contributes only weakly to the measured spectrum. When direct mechanical coupling is introduced, an additional spectral feature becomes visible. For $\eta=0.1\,\omega_m$, a third peak appears in Fig.~\ref{figure4}(b) (blue dashed curve), showing that the previously weakly participating collective motion contributes more clearly to the observed response. Increasing the coupling further to $\eta=0.15\,\omega_m$ makes this third resonance sharper and the overall peak structure more clearly resolved (red solid curve). Thus, stronger phonon-phonon coupling enhances the spectral visibility of this hybrid resonance.

Figure~\ref{figure5} presents the output-field spectrum $S_{c\mathrm{out}}(\omega)$ as a function of the normalized frequency $\omega/\omega_m$, evaluated at the same operating point as in Fig.~\ref{figure4}. For $\eta=0$, the spectrum in Fig.~\ref{figure5}(a) displays two well-resolved resonant features, again indicating that the measured optical response is dominated by two hybrid modes while the third contributes only weakly under these conditions. When the phonon-phonon coupling is introduced, this weakly visible contribution becomes more apparent in the optical output. For $\eta=0.1\,\omega_m$, a third peak emerges in Fig.~\ref{figure5}(b) (blue dashed curve), and increasing the coupling further to $\eta=0.15\,\omega_m$ enhances this feature, making the third peak sharper and easier to distinguish (red solid curve). Compared with the position-fluctuation spectra in Fig.~\ref{figure4}, the corresponding optical output spectra in Fig.~\ref{figure5} are plotted on a larger scale, so the splitting pattern is visually more prominent in the output channel.

Figure~\ref{figure6} shows the output spectrum of the $x$ quadrature, $S_{x\mathrm{out}}(\omega)$, plotted versus the normalized frequency $\omega/\omega_m$ at the same fixed operating point. In Fig.~\ref{figure6}(a), for $\eta=0$, the spectrum exhibits two distinct resonant peaks, consistent with the dominant optomechanical hybridization in the absence of direct mechanical-mechanical interaction. When the phonon-phonon coupling is introduced, the weakly participating resonance becomes resolved in this readout channel as well. For $\eta=0.1\,\omega_m$, a third peak appears in Fig.~\ref{figure6}(b) (blue dashed curve), and it becomes sharper and easier to resolve as the coupling is increased to $\eta=0.15\,\omega_m$ (red solid curve). In addition, the quadrature spectrum $S_{x\mathrm{out}}(\omega)$ is shown on a larger signal scale than the corresponding intensity spectrum $S_{c\mathrm{out}}(\omega)$ in Fig.~\ref{figure5}, so the spectral features are visually emphasized in the $x$-quadrature readout.

In Fig.~\ref{figure7}, we plot the output spectrum of the phase ($y$) quadrature, $S_{y\mathrm{out}}(\omega)$, again for the same fixed gain and power. For $\eta=0$, Fig.~\ref{figure7}(a) displays two peaks, but with a much weaker overall spectral level than in the $x$-quadrature case. As in the previous spectra, direct mechanical coupling makes the weakly participating resonance more visible. In Fig.~\ref{figure7}(b), a third peak becomes identifiable for $\eta=0.1\,\omega_m$ (blue dashed curve), and for $\eta=0.15\,\omega_m$ (red solid curve) this extra resonance is more clearly resolved. Although the third peak becomes more pronounced with increasing $\eta$, the overall magnitude of $S_{y\mathrm{out}}(\omega)$ remains substantially smaller than that of the corresponding $x$-quadrature spectrum, indicating that the phase-quadrature channel carries the same modal signatures but at a lower absolute spectral level.

\section{CONCLUSIONS}\label{section:Conclusion}
In this work, we studied how direct phonon-phonon coupling between two movable mirrors modifies the normal-mode spectrum of a parametrically assisted optomechanical cavity. In the absence of this coupling, the system shows an apparent two-peak structure because one collective mechanical mode contributes only weakly to the optical response under the chosen conditions. When direct mechanical coupling is introduced, this weakly participating mode mixes more strongly with the other hybrid modes, making a third resonance clearly resolvable. We showed that this behavior appears consistently in the mirror displacement spectrum, the output-field spectrum, and the output quadrature spectra, and becomes more pronounced as the mechanical coupling strength and drive power are increased.

From an application-oriented perspective, these results indicate that direct phonon-phonon coupling can be used as a practical control parameter for enhancing the optical readout of weak collective mechanical motion. By redistributing spectral weight among the hybrid resonances, the mechanical coupling modifies the frequency-domain response without changing the basic cavity architecture. This may be useful for multimode optomechanical readout, frequency-selective transduction, and tunable multi-resonance spectral filtering, where improved visibility of a weakly participating resonance is itself a useful functionality. In addition, the emergence and sharpening of the third resonance provide a direct spectral signature of the strength of the mechanical-mechanical interaction, suggesting that the observed spectra can also be used to diagnose and control multimode hybridization in this class of systems.

Overall, our results show that direct mechanical coupling does not create an additional eigenmode in the linearized three-mode system, but it does play an important role in determining how clearly the existing hybrid modes appear in experimentally accessible spectra. In this sense, phonon-phonon coupling provides a simple and effective way to tailor spectral visibility and multimode response in two-mirror optomechanical cavities with intracavity parametric amplification.

\textbf{Data Availability Statement} This manuscript has associated data in a data repository. All data included in this paper are available upon request by contacting with the corresponding author.
\section*{Acknowledgements}
 This work was supported by the National Natural Science Foundation of China (Grant No. 12174301), the Young Investigator (Grant No. 12305027), the Natural Science Basic Research Program of Shaanxi (Program No. 2023-JC-JQ-01), and the Fundamental Research Funds for the Central Universities.
 \section*{Appendix A}
 After taking the fourier transform of Eq.~\ref{eq:e17} to Eq.~\ref{eq:e22} we get the the position fluctuation of the movable mirror $\delta Q_1$ as
 \onecolumngrid
 \begin{eqnarray}
     &\delta Q_1(\omega)\mathcal{D}(\omega)&=e^{-\text{i}\theta}(\omega_{m}^2-\eta^2\text{cos}^2\phi)(4\text{i}c_{s}^2G\chi^2\omega_m^2(\omega_m-\eta \text{cos}\phi)(\omega_m+\eta \text{cos}\phi)+4c_se^{\text{i}\theta}\chi^2\omega_m^2\Delta c_{s}^*(\omega_m^2-\eta^2\text{cos}^2\phi)-4\text{i}e^{2\text{i}\theta}\nonumber\\[6pt]&&G\chi^2\omega_m^2c_{s}^{2*}(\omega_m^2-\eta^2\text{cos}^2\phi)-\frac{1}{2}e^{\text{i}\theta}(4G^2-(\kappa-\text{i}\omega)^2-\Delta^2)(2\omega_m(\eta^2+\text{i}\omega\gamma_m+\omega^2-\omega_m^2)-\gamma_m\eta^2\text{sin}[2\phi]))\xi_1(\omega) \nonumber\\[6pt]&&+ e^{-\text{i}\theta}(\omega_m^2-\eta^2\text{cos}^2\phi)(2\text{i}c_s^2G\chi^2\omega_m^2(\eta^2-2\omega_m^2+\eta^2\text{cos}[2\phi])+2c_se^{\text{i}\theta}\chi^2\omega_m^2\Delta c_s(\eta^2-2\omega_m^2+\eta^2\text{cos}[2\phi])\nonumber\\[6pt]&&-2\text{i}e^{2\text{i}\theta}G\chi^2\omega_m^2c_{s}^{2*}(\eta^2-2\omega_m^2+\eta^2\text{cos}[2\phi])+e^{\text{i}\theta}\eta(4G^2-(\kappa-\text{i}\omega)^2-\Delta^2)((\eta^2-\text{i}\omega\gamma_m-\omega^2-\omega_m^2)\text{cos}\phi\nonumber\\[6pt]&&+(\gamma_m-2\text{i}\omega)\omega_m\text{sin}\phi))\xi_2(\omega)+2\sqrt{2}e^{-\text{i}\theta}\sqrt{\kappa}\chi\omega_m(2c_sG+e^{\text{i}\theta}(\kappa-\text{i}(\omega+\Delta))c_{s}^*)(\omega_m^2-\eta^2\text{cos}^2\phi)(\eta\text{cos}\phi\nonumber\\[6pt]&&(-\eta^2+\text{i}\omega\gamma_m+\omega^2+\omega_m^2-\gamma_m\eta\text{sin}\phi)+\omega_m(\eta^2+\text{i}\omega\gamma_m+\omega^2-\omega_m^2-\eta(\gamma_m-2\text{i}\omega)\text{sin}\phi))\delta c_{in}(\omega)+ 2\sqrt{2}\sqrt{\kappa}\nonumber\\[6pt]&&\chi\omega_m(c_s(\kappa-\text{i}(\omega+\Delta))+2Ge^{\text{i}\theta}c_s^*)(\omega_m^2-\eta^2\text{cos}^2\phi)(\eta\text{cos}\phi(-\eta^2+\text{i}\omega\gamma_m+\omega^2+\omega_m^2-\gamma_m\eta\text{sin}\phi)\nonumber\\[6pt]&&+\omega_m(\eta^2+\text{i}\omega\gamma_m+\omega^2-\omega_m^2-\eta(\gamma_m-2\text{i}\omega)\text{sin}\phi))\delta c_{in}^\dagger(-\omega)
     \label{eq:e31}
 \end{eqnarray}

Where 
\begin{eqnarray}
  &\mathcal{D}(\omega)&=\frac{1}{4}(\eta^2-2\omega_m^2+\eta^2\text{cos}[2\phi])(-((4G^2-\Delta^2-(\kappa-\text{i}\omega)^2(\gamma_m^2(\eta^2-2\omega^2)+2(\eta-\omega-\omega_m)(\eta-\omega+\omega_m)\nonumber\\[6pt]&&(\eta+\omega+\omega_m)-4\text{i}\omega\gamma_m(\eta^2-\omega^2+\omega_m^2)-\gamma_m^2\eta^2\text{cos}[2\phi]))-16\text{i}e^{-\text{i}\theta}\chi^2\omega_m^2(-((\eta^2+\omega(\text{i}\gamma_m+\omega))\omega_m)+\omega_m^3\nonumber\\[6pt]&&+\eta(\eta^2-\text{i}\omega\gamma_m-\omega^2-\omega_m^2)\text{cos}\phi)(c_s^2G-2e^{-2\text{i}\theta}Gc_s^2+c_s\Delta c_s^*(-\text{i}\text{cos}\phi+\text{sin}\phi)))\nonumber
\end{eqnarray}

Next the correlation functions of the noise sources in the frequency domain is given by

 \begin{equation}
    2\pi\delta(\omega+\Omega)= \langle\delta_{cin}(\omega)\delta c_{cin}^\dagger(-\Omega)\rangle
    \label{eq:e32}
 \end{equation}
 
 \begin{equation}
\langle\xi_i(\omega)\xi_i(\Omega)\rangle=4\pi\frac{\gamma_m}{\omega_m}\omega[1+\text{coth}(\frac{\hbar\omega}{2K_{B}T})]\delta(\omega+\Omega)
\label{eq:e33}
 \end{equation}
 
 And the spectrum of fluctuation in position of the movable mirror is given by
 
 \begin{equation}
     2\pi S_{Q1}(\omega)\delta(\omega+\Omega)=\frac{\langle\delta Q_1(\omega)\delta Q_1(\Omega)+\delta Q_1(\Omega)\delta Q_1(\omega)\rangle}{2}
     \label{eq:e34}
 \end{equation}
 
Inserting Eq.\ref{eq:e31},\ref{eq:e32} and \ref{eq:e33} in Eq.\ref{eq:e34} the final spectrum is then given by Eq.\ref{eq:e30} in the main text, where the remaining expression for the radiation pressure and noise terms are the following

\begin{eqnarray}
    &\mathcal{R}_{ad}(\omega)&=2e^{\text{i}\theta}\kappa\chi^2\omega_m^2(2c_s^2G(\text{i}\Delta+\kappa)+c_se^{\text{i}\theta}(4G^2+\Delta^2+\kappa^2+\omega^2)c_{s}^*+2Ge^{2\text{i}\theta}(-\text{i}\Delta+\kappa)c_{s}^{2*})(\eta^2-2\omega_m^2+\eta^2\text{cos}[2\phi])^2\nonumber\\[6pt]&&(-((\eta^2+\omega(\text{i}\gamma_m+\omega))\omega_m)+\omega_m^3+\eta(\gamma_m-2\text{i}\omega)\omega_m\text{sin}\phi+\eta\text{cos}\phi(\eta^2-\text{i}\gamma_m\omega-\omega^2-\omega_m^2+\gamma_m\eta\text{sin}\phi))(-((\eta^2+\nonumber\\[6pt]&&\omega(-\text{i}\gamma_m+\omega))\omega_m)+\omega_m^3+\eta(\gamma_m+2\text{i}\omega)\omega_m\text{sin}\phi+\eta\text{cos}\phi(\eta^2+\text{i}\gamma_m\omega-\omega^2-\omega_m^2+\eta\gamma_m\text{sin}\phi))\nonumber
\end{eqnarray}

\begin{eqnarray}
    &\mathcal{T}_{h1}(\omega)&=(\frac{1}{16}(\eta^2-2\omega_m^2+\eta^2\text{cos}[2\phi])^2(2(4G^2-\Delta^2-(\kappa-\text{i}\omega)^2)\omega_m(\eta^2+\text{i}\gamma_m\omega+\omega^2-\omega_m^2)+4\chi^2\omega_m^2(\eta^2+\text{i}\gamma_m\omega-2\nonumber\\[6pt]&&\omega_m^2+\eta^2\text{cos}[2\phi])(\Delta\lvert c_s\rvert^2+Gc_{s}^{2*}(-\text{i}\text{cos}\theta+\text{sin}\theta)+c_s^2G(\text{i}\text{cos}\theta+\text{sin}\theta))+\gamma_m\eta^2(-4G^2+\Delta^2+(\kappa-\text{i}\omega)^2)\text{sin}[2\phi])\nonumber\\[6pt]&&(2(4G^2-\Delta^2-(\kappa+\text{i}\omega)^2)\omega_m(\eta^2-\text{i}\gamma_m\omega+\omega^2-\omega_m^2)+4\chi^2\omega_m^2(\eta^2-2\omega_m^2+\eta^2\text{cos}[2\phi])(\Delta\lvert c_s\rvert^2+Gc_{s}^{2*}(-\text{i}\text{cos}\theta\nonumber\\[6pt]&&+\text{sin}\theta)+c_s^2G(\text{i}\text{cos}\theta+\text{sin}\theta))+\gamma_m\eta^2(-4G^2+\Delta^2+(\kappa-\text{i}\omega)^2)\text{sin}[2\phi]))(\frac{4K_{B}T\gamma_m\omega}{\hbar\omega_m-\omega\omega_m})\nonumber
\end{eqnarray}
\begin{eqnarray}
    &\mathcal{T}_{h2}(\omega)&=(\frac{1}{4}(\eta^2-2\omega_m^2+\eta^2\text{cos}[2\phi])^2(-\eta(-4G^2+\Delta^2+(\kappa+\text{i}\omega)^2)(\eta^2+\text{i}\omega\gamma_m-\omega^2-\omega_m^2)\text{cos}\phi+2\chi^2\omega_m^2(\eta^2-\nonumber\\[6pt]&&2\omega_m^2)(\Delta\lvert c_s\rvert^2+Gc_{s}^{2*}(-\text{i}\text{cos}\theta+\text{sin}\theta)+c_s^2G(\text{i}\text{cos}\theta+\text{sin}\theta))+2\eta^2\chi^2\omega_m^2\text{cos}[2\phi](\Delta\lvert c_s\rvert^2+Gc_{s}^{2*}(-\text{i}\text{cos}\theta\nonumber\\[6pt]&&+\text{sin}\theta)+c_s^2G(\text{i}\text{cos}\theta+\text{sin}\theta))-\eta(-4G^2+\Delta^2+(\kappa+\text{i}\omega)^2)(\gamma_m+2\text{i}\omega)\omega_m\text{sin}\phi)(2c_s\Delta\chi^2\omega_m^2c_s^*(\eta^2-2\nonumber\\[6pt]&&\omega_m^2+\eta^2\text{cos}[2\phi])+2\text{i}c_s^2G\chi^2\omega_m^2\text{cos}\theta(\eta^2-2\omega_m^2+\eta^2\text{cos}[2\phi])-4c_s^2G\chi^2\omega_m^2(\omega_m-\eta\text{cos}\phi)(\omega_m+\eta\text{cos}\phi)\nonumber\\[6pt]&&\text{sin}\phi+2G\chi^2\omega_m^2c_{s}^{2*}(\eta^2-2\omega_m^2+\eta^2\text{cos}[2\phi])(-\text{i}\text{cos}\theta+\text{sin}\theta)+\eta(4G^2-\Delta^2-(\kappa-\text{i}\omega)^2)((\eta^2-\text{i}\gamma_m\omega\nonumber\\[6pt]&&-\omega^2-\omega_m^2)\text{cos}\phi+(\gamma_m-2\text{i}\omega)\omega_m\text{sin}\phi)))(\frac{4K_{B}T\gamma_m\omega}{\hbar\omega_m-\omega\omega_m})\nonumber
\end{eqnarray}
\section*{Appendix B}
After Fourier transforming Eq.\ref{eq:e17} to \ref{eq:e22} we obtained the fluctuation $\delta c(\omega)$ and by using the input-output relation $c_{\text{out}}(\omega)=\sqrt{2\kappa}c(\omega)-c_{in}(\omega)$ the output field and the quadrature fluctuations are given by

\begin{eqnarray}
    &\delta c_{out}(\omega)&=\mathcal{V}_1(\omega)\xi_1(\omega)+\mathcal{V}_2(\omega)\xi_2(\omega)+\mathcal{J}(\omega)\delta c_{in}(\omega)+\mathcal{F}(\omega)\delta c_{in}^\dagger(-\omega)
\end{eqnarray}
Where
\begin{eqnarray}
    &\mathcal{V}_1(\omega)&=\frac{1}{\mathcal{D}(\omega)}(-\text{i}\sqrt{2}e^{\text{i}\theta}G\sqrt{\kappa}\chi\omega_mc_s^*(\eta^2-2\omega_m^2+\eta^2\text{cos}[2\phi])(-((\eta^2+\omega(\text{i}\gamma_m+\omega))\omega_m)+\omega_m^3-\eta(\gamma_m-2\text{i}\omega)\nonumber\\[6pt]&&\omega_m\text{sin}\phi+\eta\text{cos}\phi(\eta^2-\text{i}\omega\gamma_m-\omega^2-\omega_m^2+\gamma_m\eta\text{sin}\phi)))(\frac{2\gamma_m(\frac{2K_{B}T}{\hbar}-\omega)}{\omega_m})\nonumber
\end{eqnarray}
\begin{eqnarray}
  &\mathcal{V}_2(\omega)&= \frac{1}{\mathcal{D}(\omega)}(-\text{i}\sqrt{2}G\sqrt{\kappa}e^{\text{i}\theta}\chi\omega_mc_s^*(\eta^2-2\omega_m^2+\eta^2\text{cos}[2\phi])(-((\eta^2+\omega(\text{i}\gamma_m+\omega))\omega_m)+\omega_m^3+\eta(\gamma_m-2\text{i}\omega)\nonumber\\[6pt]&&\omega_m\text{sin}\phi-\eta\text{cos}\phi(-\eta^2+\text{i}\omega\gamma_m+\omega^2+\omega_m^2+\gamma_m\eta\text{sin}\phi)))(\frac{2\gamma_m(\frac{2K_{B}T}{\hbar}-\omega)}{\omega_m})\nonumber 
\end{eqnarray}
\begin{eqnarray}
    &\mathcal{J}(\omega)&=\frac{1}{\mathcal{D}(\omega)}(\frac{1}{-4G^2+\Delta^2+(\kappa-\text{i}\omega)^2}\kappa(\kappa-\text{i}(\Delta+\omega)+4G\chi^2\omega_m^2c_s^*(2\text{i}c_sG+e^{\text{i}\theta}(\Delta+\text{i}\kappa+\omega)c_s)(\omega_m(\eta^2+\nonumber\\[6pt]&&\text{i}\gamma_m\omega+\omega^2-\omega_m^2)+\eta(-\eta^2+\text{i}\omega\gamma_m+\omega^2+\omega_m^2)\text{cos}\phi)(\eta^2-2\omega_m^2+\eta^2\text{cos}[2\phi]))\nonumber
\end{eqnarray}
\begin{eqnarray}
    &\mathcal{F}(\omega)&=\frac{1}{\mathcal{D}(\omega)}(\frac{1}{-4G^2+\Delta^2+(\kappa-\text{i}\omega)^2}4e^{\text{i}\theta}G\kappa(1+2\chi^2\omega_m^2c_s^*(c_s(\Delta-\text{i}\kappa-\omega)-2\text{i}e^{\text{i}\theta}Gc_{s}^*)(\omega_m(\eta^2+\text{i}\omega\gamma_m+\omega^2\nonumber\\[6pt]&&-\omega_m^2)+\eta(-\eta^2+\text{i}\omega\gamma_m+\omega^2+\omega_m^2)\text{cos}\phi)(-\eta^2+2\omega_m^2-\eta^2\text{cos}[2\phi])))\nonumber
\end{eqnarray}
We have derived the spectrum for $S_{\text{cout}}(\omega)$, $S_{\text{xout}}(\omega)$ and $S_{\text{yout}}(\omega)$ by the following relations.
\begin{eqnarray}
    2\pi S_{\text{cout}}(\omega)\delta(\omega+\Omega)=\langle\delta c_{out}^\dagger(-\Omega)\delta c_{out}(\omega)\rangle
\end{eqnarray}
\begin{eqnarray}
    2\pi S_{\text{xout}}(\omega)\delta(\omega+\Omega)=\langle\delta x_{out}(\Omega)\delta x_{out}(\omega)\rangle
\end{eqnarray}
\begin{eqnarray}
    2\pi S_{\text{yout}}(\omega)\delta(\omega+\Omega)=\langle\delta y_{out}(\Omega)\delta y_{out}(\omega)\rangle
\end{eqnarray}
\twocolumngrid

\bibliographystyle{apsrev4-2}
\bibliography{biblio.bib}
\end{document}